\documentclass[mathleft
]{an}
\usepackage{graphicx}
\usepackage{times}
\usepackage{epstopdf}
\usepackage{color}

\newcommand{\kms}{km s$^{-1}$}

\begin{document}

\Pagespan{789}{}
\Yearpublication{2006}%
\Yearsubmission{2005}%
\Month{11}%
\Volume{999}%
\Issue{88}%

\title{Constraints of a pulsation frequency on stellar parameters 
in the eclipsing spectroscopic binary system: V577 Oph}
\author{O.L. Creevey\inst{1,2}\fnmsep\thanks{Corresponding author:
  \email{orlagh@iac.es}\newline},
J. Telting\inst{3},
J.A. Belmonte\inst{1,2},
T.M. Brown\inst{4},
G. Handler\inst{5},
T.S. Metcalfe\inst{6},
F. Pinheiro\inst{7,8},
S. Sousa\inst{9},
D. Terrell\inst{10},
\and
A. Zhou\inst{11}
}
\titlerunning{Constraints of a pulsation frequency in V577Oph}
\authorrunning{Creevey et al.}
\institute{
Instituto de Astrof\'isica de Canarias, E-38200 La Laguna, Tenerife, Spain
\and
Dept. de Astrof\'isica, Universidad de La Laguna, E-38205 La Laguna, Tenerife, Spain
\and
Nordic Optical Telescope,
                Apartado 474, 38700 Santa Cruz de La Palma, Spain
\and
LCOGT Network Inc.,
6740 Cortona Dr. Suite 102, Santa Barbara, CA 93117, USA
\and 
Institut f\"ur Astronomie, University of Vienna, T\"urkenschanzstrasse 17,
A-1180 Vienna, Austria
\and
High Altitude Observatory/National Center for Atmospheric Research, 3080 Center Green,
Boulder, CO 80301, USA
\and
Centro de F\'{i}sica computacional, Universidade de Coimbra, Portugal
\and
LESIA, Observatoire de Paris, France
\and
Centro de Astrof\'isica da Universidade do Porto,
Rua das Estrelas,
4150-762 Porto, Portugal
\and
Department of Space Studies, Southwest Research Institute, Boulder, CO, USA
\and
Natl. Astronomical Observatories, Chinese Academy of Sciences, 20A Datun Road, Chaoyang District, Beijing, China}

\received{30 March 2010}
\accepted{1 April 2010}
\publonline{later}


\abstract{%
We present a preliminary spectroscopic analysis of the binary system V577Oph, 
observed during the summer of 2007 on the 2.6m NOT telescope on La Palma. 
We have obtained time series spectroscopic observations, 
which show clear binary motion as well as radial velocity variations 
due to pulsation in the primary star. 
By modelling the radial velocities we determine a full orbital solution of the 
system, which yields
    $M_A \sin^3 i = 1.562 \pm 0.012$ M$_{\odot}$ and
     $M_B \sin^3 i = 1.461 \pm  0.020$ M$_{\odot}$.
An estimate of inclination from photometry yields a primary mass of $\sim1.6$ $M_{\odot}$.
Using this derived mass, and the known pulsation frequency we can impose a lower
limit of 1 Gyr on the age of the system, and constrain the parameters of the oscillation mode.  
We show that with further analysis of the spectra (extracting the atmospheric parameters),
tighter constraints could be imposed on the age, metallicity and the mode parameters.
This work emphasizes the power that a single pulsation frequency 
can have for constraining stellar parameters in an eclipsing binary system.}

\maketitle

\section{Introduction}

{Large amplitude} $\delta$ Scuti stars are 1.5 - 2.5 M$_{\odot}$ mostly MS stars, with few observed
pulsation modes.
Interpreting these pulsations 
is difficult, mainly because
1) generally, the fundamental parameters of the star are poorly constrained, and 2)
we do not know enough information about the mode characteristics 
(degree $l$, azimuthal $m$, and radial order $n$) to allow us to use the mode
as an extra observable constraint.
To overcome these obstacles we can 1) study a pulsating star in a detached binary
system, where the fundamental
parameters can be determined
 independent of using the pulsation 
mode, and 2) use new techniques to study the observational 
profiles of the oscillations to constrain its mode properties.

V577 Oph is a detached eclipsing binary system with a
$\delta$ Scuti component, and is 
an ideal candidate for a seismological analysis.
Its orbit is eccentric ($e=0.22$), the dominant pulsation {period} of the 
$\delta$ Scuti object is of the order of 
1.3 hours, 1/100th the size of its orbital period, and the few published 
light curves indicate no out-of-eclipse modulations.  
All of this indicates that the system is detached, and so we may 
assume
that the $\delta$ Scuti evolves as a single star i.e. no mass transfer, tidal distortions or
oscillations produced by a tidally-locked orbit.  

The system was first studied by \cite{vol90}.
The author concluded that 
the primary component (the more massive and more luminous) is the pulsating 
star, due to the amplitude changes during primary and secondary eclipse.
They also measured B-V = 0.488 mag outside of eclipse, 0.504 mag during primary eclipse minimum, 
and 0.509 mag during
secondary eclipse minimum.
Using the photometric eclipse depths, the orbital period $\Pi$ has
 been measured as  6.079 days, and
a dominant pulsation frequency of 0.0695
days has been detected. 
The spectral types of both components have been estimated as A8 from 
interstellar redenning (\cite{shu85} --- referenced in Diethelm 1993),
and more recently, Zhou (2001) obtained new photometric light curves
and extracted the dominant pulsation frequency of 
14.390 cycles per day 
with an oscillation 
amplitude of 0.029 mag.

In this paper, we discuss our recent spectroscopic observations of V577Oph.  
As an initial analysis, we extract the 
the radial velocities and determine the orbital parameters.
Using an estimate of inclination from photometry to obtain the 
mass, we use seismic
models to discuss the global properties of V577 Oph.


\section{Observations}

We obtained 5 consecutive nights on the Nordic Optical Telescope (NOT),
located in the Observatorio del Roque de los Muchachos on La Palma, 
Canary Islands. 
The NOT is a 2.6m telescope with an alt-azimuth mount.
The instrument used is the FIbre-fed Echelle Spectrograph (FIES) ---
a cross-dispersed high-resolution echelle spectrograph with a maximum
resolution of R=65,000.  Our setup was medium-resolution with R=45,000, with 
spectral range: 3640 - 7455.  The maximum system
efficiency (detector-instrument-fiber-telescope-atmosphere at airmass 1.0) 
is 9\% at 6000 \AA. 

During the nights of July 4 - July 8, 2007 inclusive, we obtained 
spectra of V577 Oph, using 10 minute exposures. Accounting for calibration, 
{\it Targets of Opportunities} and other interruptions, each night 
an average of 32 science images in sets of 
4 for V577 Oph were taken.
Due to the faintness of our source (V = 11.0) and the potential
of overexposing ThAr,
we opted to obtain the calibration frames at regular intervals of 
about 40-45 minutes.
These frames were used to calibrate the 
wavelength of the two images before and after this frame.
This results in a loss of accuracy in wavelength, but of the order of
0.002~\AA~at $\lambda$ = 3659.6294~\AA. 
This corresponds approximately to an error of 0.1 -- 0.2 \kms.

We used the {\tt FIEStool} (Stempels, 2004)\footnote{{\tt 
http://www.not.iac.es/instruments/fies/\\fiestool/FIEStool.html}}
reduction software to calibrate the wavelength
and reduce the spectra.  
This tool is optimized for FIES data, but 
the reduction is standard and calls {\tt IRAF} to perform some of the 
tasks.
We subsequently shifted the spectra to barycentric frame.

    \begin{table*}
      \begin{center}
	\caption{Orbital Elements from Radial Velocity Curves  \label{tab:v577_rvele}}
	\begin{tabular}{llrrrrrrrrrrrr}
	  \hline\hline
  && $P_0$ & $P_{AB}$ & $P_{A}$ & $P_{B}$ & ($\sigma$) & 
  $P_{AB}$& $P_{A}$ & $P_{B}$ & ($\sigma$) & P\\
  \hline
$e$ &     & 0.2 &0.212 &0.169&0.223 &(0.004) & 0.204& 0.167&0.224  &(0.006) & 0.204\\
$\gamma$&(\kms)&-37.0&-38.38&-38.26&-37.62&(1.34) &-38.24&-38.04&-37.66 &(0.87) & -38.24\\
$\omega$ &($^{\circ}$)&47.0&50.13&46.39 & 47.37 &(0.23) & 47.80& 47.59& 47.69&(0.22)&47.80\\
$K_A$ &(\kms) &85.0&83.78 &83.53 & 83.79 &(0.60) & 83.99& 83.49& 83.99&(0.58) &83.99\\
$K_B$ &(\kms)& 90.0&89.65 &89.65 & 90.05 &(0.27) &89.79 & 89.79& 90.06&(0.35)&89.79\\
$T_0$ & (days)&2.0 &1.727 &1.669 &1.702 &(0.017) &1.699 & 1.682& 1.706 &(0.01)&1.699\\
$\Pi$ &  (days)& 6.079&6.145 &6.053 & 6.072&(0.030) &&&&& 6.079\\
$\chi^2_R$ &&&0.649&0.691&0.523 & &0.672 & 0.687&0.520\\
  \hline\hline
  \end{tabular}
	\end{center}
      \end{table*}

\section{Analysis}
\subsection{Determining the radial velocities}
\begin{figure}
      \center{\includegraphics[width = 0.4\textwidth]
	{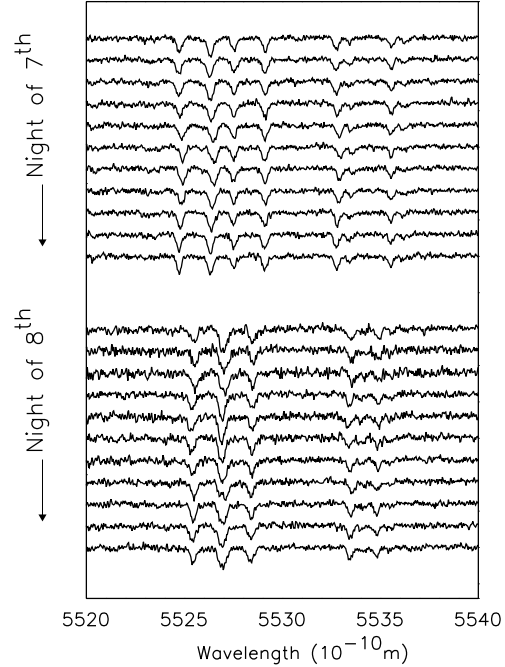}}
      \caption{Small spectral range showing the evolution of the doubled
        spectral lines.
	In some places it is clear to see the evolution of one line, while in 
	others the lines are blended especially at maximum separation.
	\label{fig:v577_selectlines}}
    \end{figure}

    Because the spectral lines are not very broad, and the binary motion can be easily
    seen, we chose
    to measure the radial velocities by  modelling a set of absorption profiles with
    Gaussian functions, where
    the central position of the Gaussian corresponds to the measured radial velocity.
    We identify the lines by eye, by inspecting the time series such
    as that shown in Fig.~\ref{fig:v577_selectlines}.
    We chose the lines that are not blended with other spectral features e.g.
    those around 5535 \AA\ in Fig.~\ref{fig:v577_selectlines}.
    The lines 
    between 5525 and 5530 \AA~were not chosen, because 
    near relative radial velocities of zero, it is difficult to 
    measure these blended lines.

    Figure \ref{fig:v577_rvfit} (top panel) shows the fitted radial velocities using the 
    spectra in a range around 5183~\AA\
    as a function of phase $\phi$ from an arbitrary reference phase.
    The fitted values are shown within the dashed lines, which represent the
    beginning and end of one orbital period. 
   At $\phi < 0$ and $\phi > 1$, the same data are plotted modulo $\Pi$ (orbital period).
    The 
    large {\it noise} associated with the primary component (positive radial velocities for 
    $\phi < 0.4$) is 
    the dominant oscillation mode with an amplitude of about 5 \kms. 
    An initial inspection of these results indicates that the components of the 
    system are of similar mass --- the spectral features also show evidence of 
    this.
     
    \subsection{Orbital solution}
    We determined the orbital solution by fitting simultaneously the free parameters:
    the semi-amplitude of the radial velocity curves for component $A$ and $B$, 
    $K_A$ and  $K_B$, the systemic velocity $\gamma$, the eccentricity $e$ with 
    argument of periastron $\omega$, as a function of either time $t_i$ or phase $\phi_i$,
    where $\phi_i = \frac{t_i \mbox{\small{mod}} {\Pi} }{\Pi } - T_0$ and
    $T_0 = 0$ is an arbitrary reference value.
    When we fitted as a function of phase, the reference phase $\phi_0$ 
    was fitted with $\Pi$ fixed, and when
    fitting with time we fitted both $T_0$ and $\Pi$.

    We fitted the data using a $\chi^2$ test:
    \begin{equation}
    \chi^2 = \sum_{i=1}^N \left (\frac{o_i - M_i}{\epsilon_i} \right)^2,
    \label{eqn:chisq}
    \end{equation}
    where $o_i$, $\epsilon_i$ and $M_i$ are respectively the $N$ observations, measurement 
    errors and the model observables produced from a specific set of model parameters.
    An initial guess of the parameters are needed, and these are obtained by 
    using the published information about the photometrically derived 
    $\Pi$ and $e$, and directly inspecting 
    the radial velocity curves.
    The initial values are shown in the first column of
    Table \ref{tab:v577_rvele}.
    The convergence criteria was given by setting $\chi^2_R$ (reduced $\chi^2$) to 1 as well 
    as testing how the free parameters changed with each iteration of the minimization algorithm.
    When the parameters were considered "stable" the minimization was usually stopped.
    
    The fitting was performed by modelling the individual radial velocity curves i.e. just one of
    $K_A$ and $K_B$ and by 
    modelling both simultaneously.  
    These are denoted respectively in Table \ref{tab:v577_rvele} as
    the column headings $P_A$, $P_B$ and $P_{AB}$.  
    Note that when modelling only one curve,
    and when the parameters converged, 
    we fixed this solution to subsequently 
    determine $K_A$ or $K_B$ from the data of the other component.
      The first set of columns show the results when we fitted $\Pi$ and $T_0$ as a function of
    time, while the second set of columns show the results when we use phase (note that
    the parameter $T_0$ has the meaning of $\phi_0$ for this second set of solutions). 
    The fourth column lists the average value of the uncertainties
    from these fits.


\section{Results}

     
    The fitted parameters and uncertainties ($\sigma_i$)  are shown 
    in Table~\ref{tab:v577_rvele} (see above paragraph for details).
    $\gamma$, $\omega$, $K_A$ and $K_B$ return stable values while 
    using all six fitting functions (data columns 2 -- 4 and 6 -- 8): 
    all values fall to within their quoted uncertainties.
    Additionally there is no qualitative difference between these 
    fitted parameters while leaving $\Pi$ free or fixing it, 
    the only variation found is the value of the uncertainty on $\gamma$.
    
\begin{figure}
      \center{\includegraphics[width=0.47\textwidth,height = 0.2\textheight]
		      {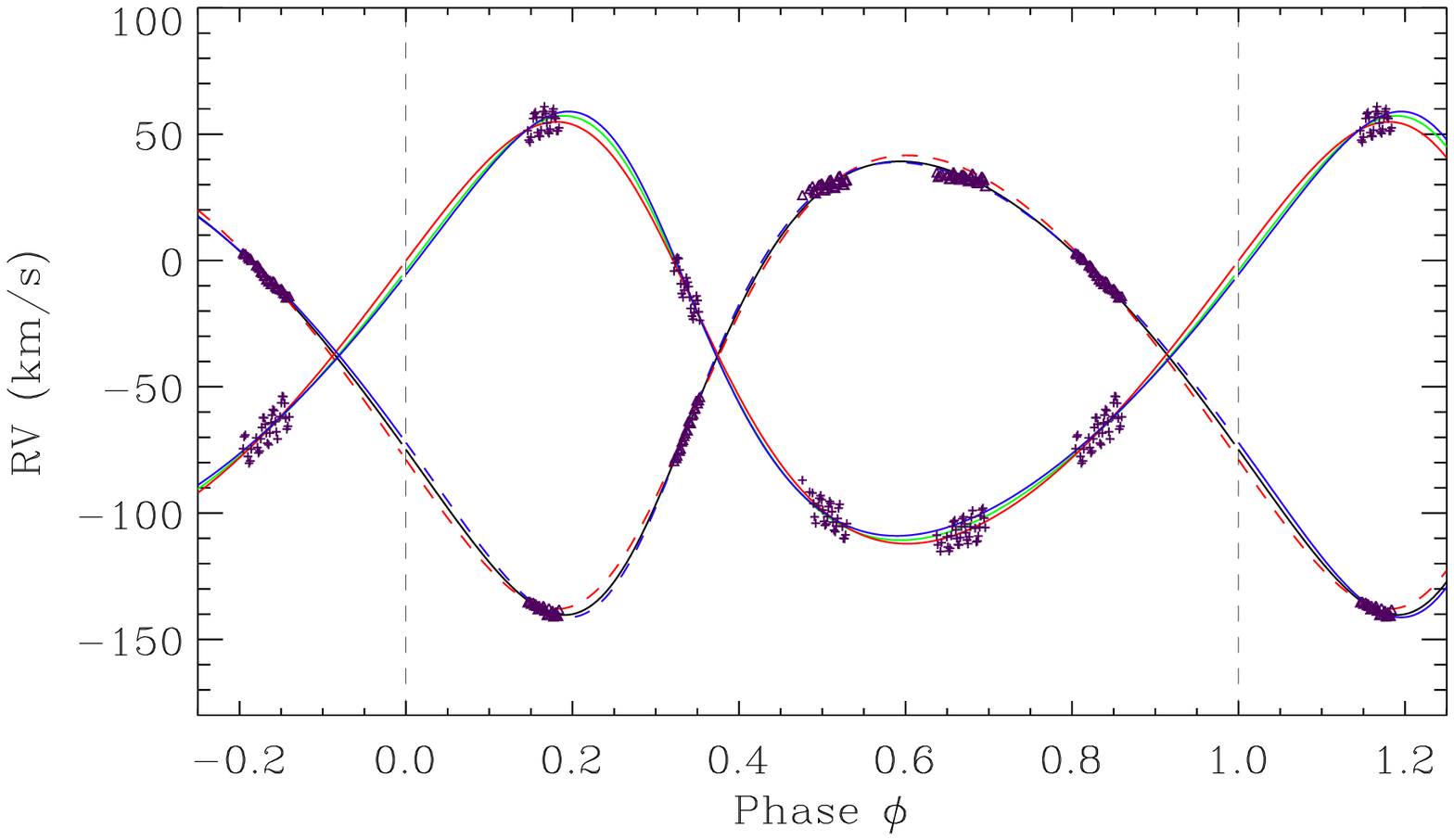}
      \includegraphics[width=0.47\textwidth,height = 0.15\textheight]
		      {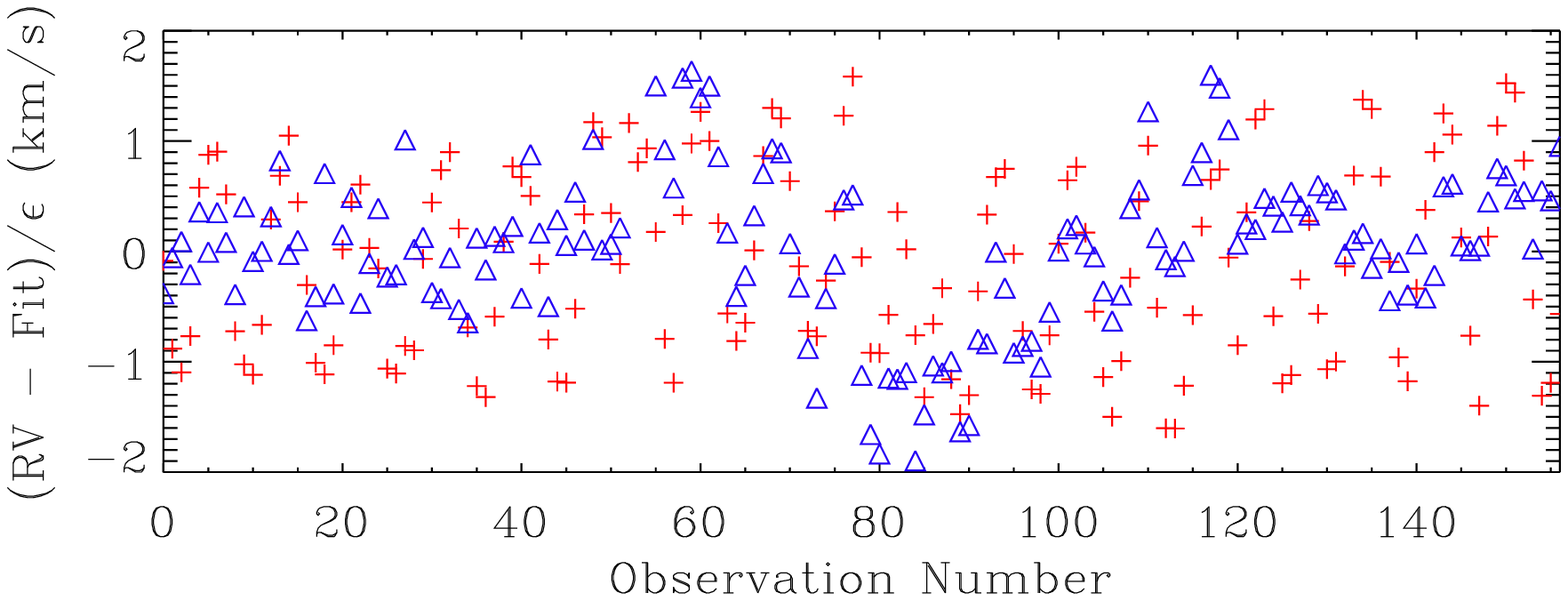}}
      \caption{
	{\sl Top:} The orbital solution from fitting the radial velocity data.
	The data within the vertical dashed lines are the true 
	data points, those at phase $\phi < 0$ and $\phi > 1$ are the data repeated modulo $\Pi$.
	The green, red and blue curves represent the resulting 
	fit curves from $P_B$, $P_A$ and $P_{AB}$.
	{\sl Bottom:} Residual of data minus fit radial velocity curve with $P_{AB}$ for 
	the primary (red) and the secondary (blue).
	\label{fig:v577_rvfit}}
    \end{figure}

    The initial value of eccentricity was estimated as $e = 0.22$ (Shugarov 1985). 
    When we fitted the radial velocities using the individual curves and then using both
    together we obtain very different values: 0.207, 0.168 and 0.224, with these
    values almost reproduced while fitting with $\Pi$ fixed.
    Figure \ref{fig:v577_rvfit} shows the fitted radial velocity curves for each of 
    the second three solutions.
    The green, red and blue curves show the orbital solutions from the fits using 
    both sets of radial velocity data, the primary component only and the secondary 
    component only, respectively.
    To account for the observed dominant 
    oscillation amplitude, 5 kms$^{-1}$ was added to the 
    measurement errors for the primary component for the fitting process.

    Obtaining different fitted values for $\Pi$ will affect the fitted value 
    of $T_0$. 
    This is the reason that a larger difference between the $T_0$
    values is seen for the first three solutions, than for the second three
    solutions.
    While keeping $\Pi$ fixed, the variation in 
    $T_0$ is only within $2\sigma$.

    Because of the large amplitude residual, it would be less feasible to
    regard the fit using the pulsating component data only as the optimal (red curve).
    Inspecting the data points (for the secondary component) 
    and the red  curve indicates that the fit is not optimal for both components.
    For this reason, the choice of $e = 0.16$ is discarded.

    The fitted values while using the data from both stars is adopted (green curve), and are
    given in Table~\ref{tab:v577_rvele} last column under "P".
    This solution is consistent with the solution while fitting just the data
    from the non-pulsating component (blue).
    Additionally, there is no difference in the derived values of 
    $a \sin i$, $M_A \sin^3 i$ and $M_B \sin^3 i$ 
    when we fit $\Pi$ and when it is fixed.
    The lower panel of Fig.~\ref{fig:v577_rvfit} shows the residual
    for the data from the primary pulsating component divided by the 
    observational errors (including the additional 5 \kms added to the observed
    error) 
    in red, and for the secondary component
    in blue.
    All of the data points fall to within 2$\epsilon$ indicating an
    adequate fit.

     Using the orbital solution given under the column heading P 
     from  Table \ref{tab:v577_rvele} we determine the system's physical parameters:
     $a \sin i$ = 0.0935 $\pm$ 0.0003 AU,
    $M_T \sin^3 i$ = 3.024 $\pm$ 0.023 $M_{\odot}$, 
    and $q$ = $M_B/M_A$ = 0.939 $\pm$ 0.006.
   Table~\ref{tab:v577_physical} summarizes the 
    component and orbital parameters.
    
   
    \begin{table}
      \begin{center}
	\caption{Physical Stellar \& Orbital Parameters\label{tab:v577_physical}}
	\begin{tabular}{lrrclrr}
	  \hline\hline
	  & $P$ & $\sigma$ \\
	  \hline
	  $M_A \sin^3 i$ (M$_{\odot}$) &1.562 & 0.012\\
	  $M_B \sin^3 i$ (M$_{\odot}$)& 1.461 & 0.020\\
	  $a \sin i$ (AU)& 0.0942 & 0.0003\\
	  $\Pi$ (days) &6.0791 & 0.0001\\
	  \hline\hline
	\end{tabular}
      \end{center}
    \end{table}

\section{Seismic inference}
\begin{figure}
    \center{      \includegraphics[width=0.45\textwidth]{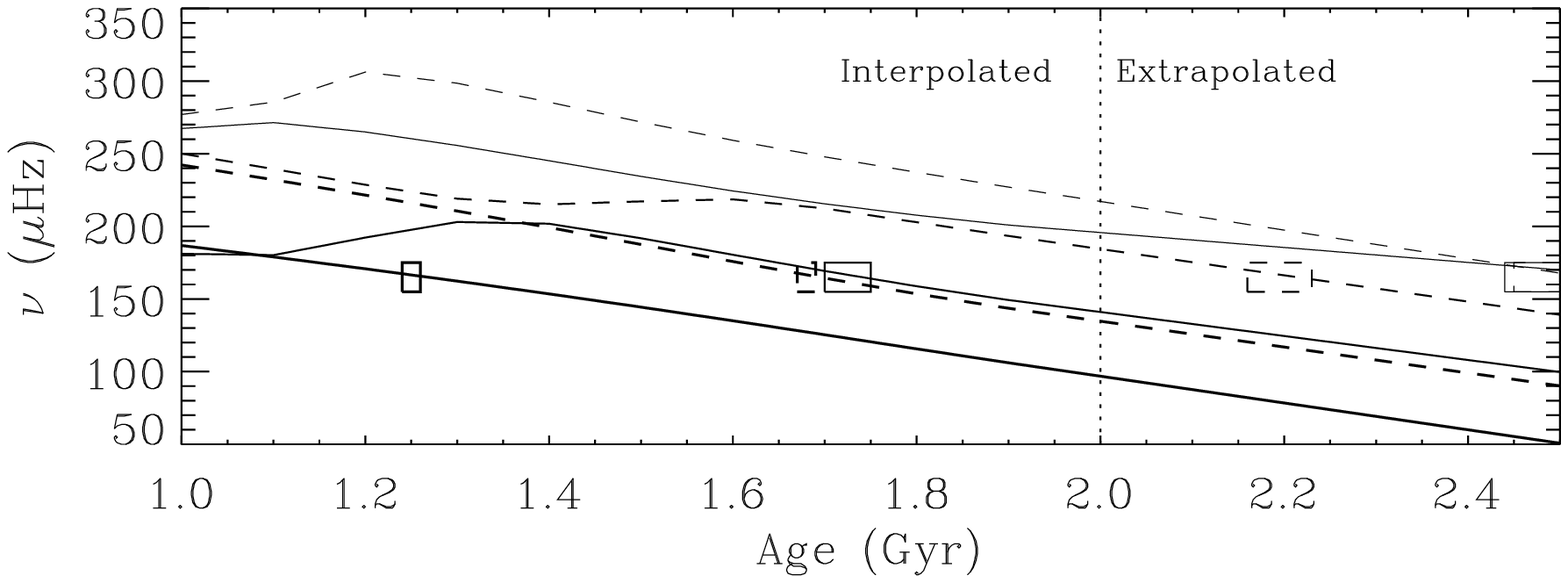}
      \includegraphics[width=0.45\textwidth]{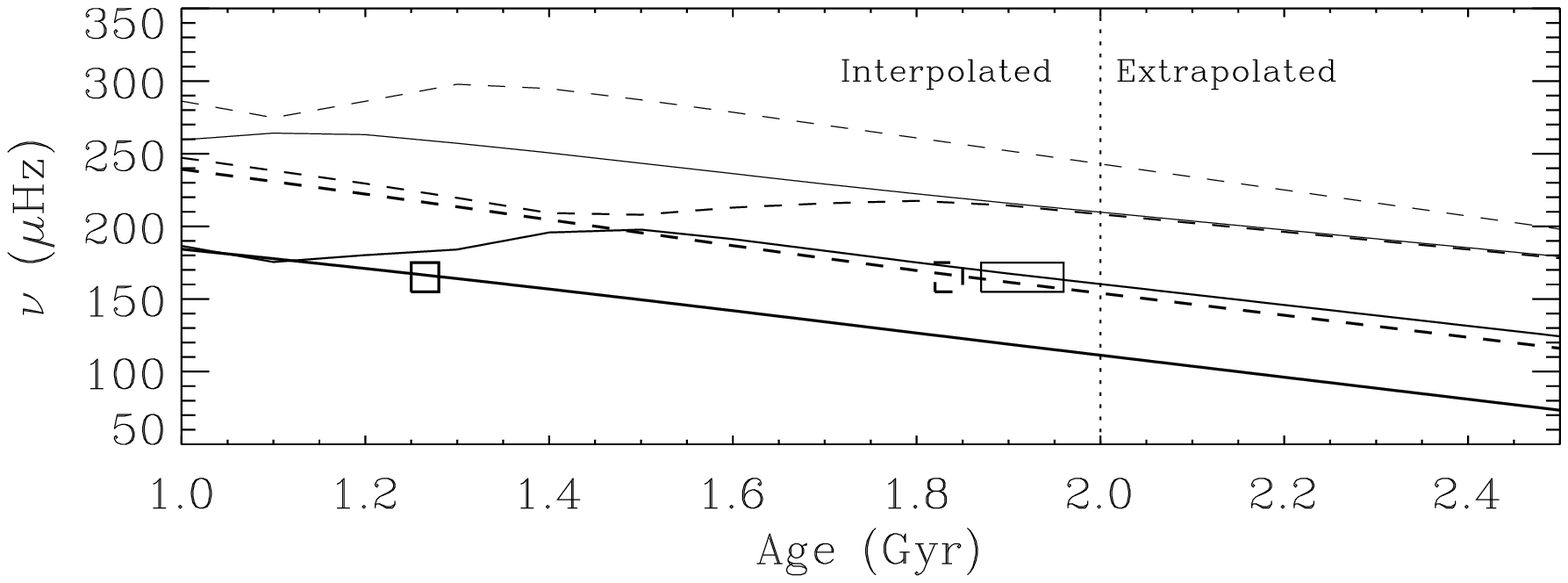}}
    \caption{Frequency values  for various low degree modes for a 
      1.63 M$_{\odot}$ star (V577 Oph A)
      as a function of age for $Z$ = 0.020 (top) and 0.025 (bottom). See text for details.
      \label{nuage}}
\end{figure}

An estimate of the inclination of the system from photometry ($i>80^{\circ}$) 
yields a primary mass of $\sim1.6$ M$_{\odot}$ (Zhou, priv. comm.).
Figure~\ref{nuage} show theoretical frequencies of a 1.63 M$_{\odot}$ 
star for 2 values of metallicity: $Z$ (initial metal mass fraction) = 0.020 and 0.025.
These mode frequencies were calculated using the ASTEC and ADIPLS stellar structure, evolution and 
pulsation codes (Christensen-Dalsgaard 2008a,b).
The lines correspond to {fundamental (continuous) and first overtone (dashed)}
oscillation modes of different degree ($\l = 0,1,2$ from {bottom to top}).
The boxes represent the observed value of the frequency which includes an error
to take into account that the frequency could have a non-zero azimuthal order ($m$ where $-l < m < l$).

{If we assume that no g-modes are present, since these are not common among 
large amplitude $\delta$ Scuti stars}, 
Figure 3 {then} shows that the system must have a lower age 
limit of 1 Gyr (younger stars exhibit higher frequencies). 
If we knew which mode  was observed we
could further constrain the age to either 1.2 -- 1.3 Gyr or 1.6 -- 2.0 Gyr.
Also we see that there are three possible solutions to the identity of the mode, without 
using any  observational technique to do this.  
{These possibilities are: radial mode (fundamental or first overtone) or fundamental 
non-radial mode $\ell = 1$.}
We can also see that a metallicity measurement and a mode-identification
would provide further constraints on the age and chemical composition.

\section{Conclusions}
We obtained 5 nights of spectroscopic observations of the eclipsing binary with a $\delta$ Scuti component V577 Oph.
We extracted the radial velocities of both components, which shows very clear 
orbital motion of both stars and pulsations in the primary (more massive) star.
We fitted the velocities to solve for the orbital parameters
and then derive for the components   $M_A \sin^3 i = 1.562 \pm 0.012$ M$_{\odot}$ and
     $M_B \sin^3 i = 1.461 \pm  0.020$ M$_{\odot}$.
An estimate of inclination implies a primary pulsating component of $\sim1.6$M$_{\odot}$.



A dominant oscillation frequency has been confirmed, and comparing this with some stellar
evolution and pulsation models allows us to impose a lower limit of 1 Gyr on the age of the star,
and constrains the identification of the mode to either
{radial mode (fundamental or first overtone) or fundamental 
non-radial mode $\ell = 1$.}
We plan to disentangle the spectra in order to determine the 
atmospheric parameters of $T_{\rm eff}$, [M/H], and $\log g$. 
This information would additionally constrain both the 
age of the star and the pulsation mode parameters.

\acknowledgements
This data has been collected using the NOT telescope operated at the 
Observatorio del Roque de los Muchachos 
on the island of La Palma, Canary Islands.
This work was partially funded by a Newkirk Graduate Research grant at 
the High Altitude Observatory, Boulder, Colorado, USA.
F.J.G.P. acknowledges grant /SFRH/BPD/37491/2007 from F.C.T.


\end{document}